\documentstyle[eqsecnum,aps,prd,epsf]{revtex}

\def\sqr#1#2{{\vcenter{\hrule height.#2pt
      \hbox{\vrule width.#2pt height#1pt \kern#1pt
          \vrule width.#2pt}
      \hrule height.#2pt}}}

\def\edth{{\rlap{$\partial$}\raise0.3em\hbox{$-$}}}

\def\gtrless{{\hbox{\raisebox{0.6ex}{$\,>$}}\kern-1.8ex
                   \hbox{\raisebox{-0.6ex}{$<\,$}}}}
\def\msol{\ifmmode M_\odot\else$M_\odot$\fi}

\begin{document}


%

\bigskip
\bigskip
\bigskip

\begin{center}

{\Large \bf
Adiabatic Expansion for Metric Perturbation 
and the condition to solve the Gauge Problem 
for Gravitational Radiation Reaction Problem
}\\
\bigskip

{\large Yasushi Mino
\footnote{Electronic address: mino@tapir.caltech.edu}}\\

\medskip

{\em mail code 130-33 California Institute of Technology Pasadena CA 91125 USA}\\ 

\medskip

\today

\end{center}

\bigskip

{\bf Abstract} \\
We examine the adiabatic approximation 
in the study of a relativistic two-body problem 
with the gravitational radiation reaction. 
We recently pointed out that the usual metric perturbation scheme 
using a perturbation of the stress-energy tensor 
may not be appropriate for study 
of the dissipative dynamics of the bodies due to the radiation reaction. 
Over a time scale during which the usual perturbation scheme is valid, 
the orbits may not deviate substantially 
relative to the orbits of the background orbits. 
As a result, one can eliminate the orbital deviation 
through a gauge transformation. 
This is called the gauge problem 
of the gravitational radiation reaction exerted on the bodies, 
and it has been reported that a careful gauge fixing may be necessary 
to produce a physically reasonable prediction 
for the evolution of the system.  

We recently proposed a possible approach to solve this problem 
with a linear black hole perturbation. 
This paper proposes a non-linear generalization of that method 
for a general application of this problem. 
We show that, under a specific gauge condition, 
the method actually allows us to avoid the gauge problem.

\section{Introduction} \label{sec:intro}

Many-body problems are fundamental problems in general relativity 
and have a long history of theoretical investigation. 
Starting from the famous paper of Einstein, Infeld and Hoffman,\cite{EIH} 
the equation of motion of point particles was studied by various authors. 
The pioneering work of Einstein, Infeld and Hofman\cite{EIH} 
used Dirac's delta functions as particles 
(the point particle approximation), 
and assumed a coordinate system 
in which the spacetime is weakly curved 
and the motion of the particles is sufficiently smaller 
than the light speed (the post-Newtonian approximation). 
To leading order in this approximation, 
we obtain a Newtonian equation of motion. 
Aside from the point particle approximation, 
one can perturbatively consider 
higher-order terms of the post-Newtonian approximation, 
and Einstein, Infeld and Hofman derived the equation of motion 
with the leading relativistic effect 
under the post-Newtonian approximation. 

Much effort has been dedicated to deriving an equation of motion 
for a two-body system in a higher order of this approximation.\cite{pn} 
Such attempts were motivated by recent ground-based gravitational-wave 
interferometric detection projects, 
such as LIGO, VIRGO, TAMA and GEO.\cite{GW} 
These detectors are expected to facilitate 
observation of gravitational waves from a close binary system 
consisting of stellar mass compact objects 
and, in order to extract astrophysical information 
concerning the gravitational-wave source from the observed waves, 
it is necessary to possess a theoretical gravitational waveform 
from a binary system. 
Such a binary is called an equal mass binary system 
and is believed to be accurately approximated 
by a relativistic two-body system, 
because the dynamical effect caused by the environment 
may not be important. 
Using the gravitational-wave generation formalism,\cite{pn} 
one can calculate the gravitational waveform from a two-body system, 
whose dynamics are governed by the relativistic equation of motion. 
We expect that detected gravitational waves might reveal 
a fundamental property of the gravitational law 
in the strongly gravitating limit.\cite{GW} 

The study of the gravitational waveform for a binary 
has also been carried out using black hole perturbations.\cite{mp} 
In this case, we consider the case that 
the mass ratio of the binary is extreme, 
such as in the case of a binary consisting of a supermassive black hole 
and an inspiralling stellar mass compact object. 
It is expected that gravitational waves from such a binary system 
will be detected by a space-based gravitational-wave 
interferometric detector, such as LISA.\cite{GW} 
Because of the extreme mass ratio, 
the gravitational potential of the system is 
due almost entirely to the heavier body, 
and we may suppose that the heavier body creates a background geometry, 
which is assumed to be a black hole geometry. 
The gravitational effect caused by the lighter body is treated 
as a metric perturbation of the background black hole geometry. 
Given the motion of the lighter body, 
the black hole perturbation formalism yields 
the gravitational waveform from the binary system,\cite{mp} 
with which we may study the gravitational theory. 

During the observation time in both cases, 
the gravitational radiation reaction becomes important 
for the orbital motion of the binary. 
In this paper, we discuss the problem of the mathematical framework 
for the study of this effect. 
Until this time, in most cases, 
the gravitational radiation reaction effect 
has been dealt under the so-called adiabatic approximation. 
There are two key assumptions for the adiabatic approximation. 
The first approximation is that 
the instantaneous dynamics of the orbit are conservative, 
because the radiation reaction effect is weak. 
We specifically consider the case in which an orbit is characterized 
by several constants of motion, denoted by ${\cal E}^a$, 
and we consider the evolution of such ``constants'' of motion 
instead of the orbital coordinates of the bodies. 
With this approximation, we calculate 
the radiation reaction exerted on the ``constants'' of motion 
by the conservative orbit of given constants of motion. 
Thus, we have an evolution equation of the form 
\begin{eqnarray}
\frac{d{\cal E}^a}{dt} = {\cal F}^a({\cal E}) \,. \label{eq:radf} 
\end{eqnarray}
The second assumption is that 
the time averaged change of the ``constants'' of motion 
$\large<d{\cal E}^a/dt\large>$ 
describes the orbital evolution, 
because the radiation reaction affects 
the orbital dynamics in a secular manner. 
The ``constants'' of motion evolve 
in the so-called radiation reaction time ($T_{\hbox{rad}}$), 
and it is believed that the adiabatic approximation holds 
when the radiation reaction time is much longer 
than the dynamical time scale of the orbit ($T_{\hbox{dyn}}$). 

In the case of an equal mass binary, 
we consider a nonlinear metric perturbation 
with a flat metric as the background.\cite{pn} 
We separate the metric perturbation 
into the symmetric and anti-symmetric parts 
with respect to the background coordinate time. 
The conservative motion of the bodies 
is derived using the symmetric part of the metric, 
and it is derived simultaneously and consistently 
with the symmetric part of the metric 
under the post-Newtonian expansion.\cite{pn} 
When the spins of the bodies are negligible, 
the conservative motion of the binary is characterized 
by two constants of motion, the orbital energy and the angular momentum. 
Using the conservative motion of the bodies as a source, 
one can derive the anti-symmetric part of the metric 
with the wave generation formalism, 
employing the post-Minkowskian expansion.\cite{pn} 
This part is responsible for the gravitational radiation reaction 
and, by evaluating the gauge-invariant flux 
of the gravitational radiation momentum escaping to infinity,\cite{flx} 
we obtain the time averaged losses 
of the orbital energy and the angular momentum 
as functions of the constants of motion. 

In the case of the extreme mass ratio binary, 
we take the evolution of a geodesic of the background black hole geometry 
as a conservative motion of the orbit. 
When the background is a Kerr black hole, 
the geodesic is characterized by three ``constants'' of motion, 
the orbital energy $E$, 
the $z$-component of the angular momentum $L_z$, 
and the Carter constant $C$. 
The metric perturbation induced by the geodesic is responsible 
for the gravitational radiation reaction. 
The radiation reaction exerted on these ``constants'' 
is derived by evaluating the radiative part of the self-force, 
\footnote{It is believed that the self-force has a conservative component. 
However, in Ref.\cite{adi}, we find that one can eliminate 
this part by taking the radiation reaction gauge condition 
over the radiation reaction time. 
This shows that 
the conservative component of the self-force is not observable 
unless we can observe gravitational waveforms 
with a precision of the mass ratio in each period.}
and we derive a formula 
for the time averaged losses of these three ``constants'' 
as functions of the constants of motion. 
It is also shown that these time averaged losses are gauge invariant 
in Ref.\cite{rad}. 

Thus, we have successfully formulated 
the radiation reaction problem in both cases 
under the adiabatic approximation, 
and the results in both cases are shown to be consistent 
in the post-Newtonian limit 
and in the extreme mass ratio limit.\cite{pn,mp} 

The adiabatic approximation is merely an approximation 
for a practical calculation of the gravitational radiation reaction effect, 
and it would eventually be invalid for an accurate prediction 
of gravitational waveforms. 
In this paper, we examine the validity of the adiabatic approximation 
by estimating the effect on the orbital evolution 
which appears when the adiabatic approximation becomes invalid. 
Regarding the second assumption, 
an explicit calculation of the part of the self-force 
other than the time averaged losses of the ``constants'' 
was done in the equal mass binary case.\cite{phase} 
In the extreme mass ratio binary case, 
we showed that the time averaged losses of the ``constants'' 
are the only gauge invariant components of the self-force, 
and that one may take the gauge transformation 
to eliminate the gauge dependent components 
on the radiation reaction time scale. 
In this paper, we study 
the first assumption of the adiabatic approximation. 
To this point, we have discussed the adiabatic approximation of the orbit. 
In Ref.\cite{adi}, we pointed out that 
for a consistent study of the validity of the adiabatic approximation, 
it is necessary to consider the adiabatic expansion of the metric 
together with the adiabatic evolution of the orbit. 
We carried out a preliminary investigation of the adiabatic expansion 
of the metric in Ref.\cite{adi} 
and here we extend the idea proposed there. 

The organization of this paper is as follows. 
In \S \ref{sec:ada}, we briefly summarize our present result 
concerning the validity of the adiabatic approximation. 
We have previously carried out several investigations 
of the validity of the second assumption. 
However, the validity of the first assumption 
has not been carefully studied until now. 
In \S \ref{sec:pri}, 
we argue that the question concerning the first assumption 
is related to the basic mathematical scheme 
used to treat a binary system coupled to a gravitational field. 
In \S \ref{sec:rmp} and \ref{sec:amp}, we propose 
a new metric perturbation scheme 
and its practical calculational procedure 
appropriate for studying the evolution of the metric and the orbit. 
In \S \ref{sec:bur}, we show that the resulting perturbation expansion 
of the metric behaves properly under a specific gauge condition 
on the second-order metric perturbation. 
We summarize results in \S \ref{sec:con}.

\section{Adiabatic approximation} \label{sec:ada}

As pointed out in the previous section, 
the adiabatic approximation uses two separate assumptions. 
Here we summarize our present result 
on the validity of these two assumptions separately. 

The first assumption is to use the conservative orbit 
as a source to calculate the radiative component 
of the gravitational field. 
The conservative orbit is believed to be a good approximation 
of the orbit instantaneously, 
but, the orbit will deviate from a conservative orbit 
due to the radiation reaction. 
In order to examine its validity, it is necessary to consider 
the component of the gravitational field induced by the orbit 
which evolves due to the radiation reaction. 
At this point, it is important to note that 
the orbital coordinates without the radiation reaction 
are characterized by not only the constants of motion, ${\cal E}$, 
but also the so-called phase constants, ${\cal C}$ 
(see \S \ref{sec:pri}). 
While the effect of the evolving ``constants'' of motion 
on the metric perturbation has been studied, 
the effect of the evolving phase ``constants'' 
had not been carefully considered 
until we pointed it out in Ref.\cite{adi}. 
This is because one can set the phase constants to zero 
for a conserved orbit through an appropriate isometry transformation, 
but once we consider the orbital evolution 
due to the radiation reaction, 
one cannot set the phase ``constants'' to zero during the evolution. 
Because the evolution time scale for the phase ``constants'' 
is shorter than that of the ``constants'' of motion, 
it is more important to consider 
the evolution of the phase ``constants''.\cite{adi} 

A consistent calculation of such a metric can be done 
by calculating the evolution of the metric and the orbit together. 
In Ref.\cite{adi}, we propose 
an adiabatic approximation for both the metric and the orbit 
in the case of an extreme mass ratio binary system, 
and we find that one can predict the orbit 
under this first assumption over the radiation reaction time scale. 

Because the radiation reaction time scale is 
approximately equal to or less than the observation time scale 
in the case of an equal mass binary 
for an observation by ground-based detectors 
and in the case of an extreme mass ratio binary 
for an observation by a space-based detector, 
the result given in Ref.\cite{adi} may pose a serious question 
concerning the reliability of theoretical waveforms 
derived using the adiabatic approximation. 
In \S \ref{sec:pri}, we discuss the consequences 
of this problem in greater detail, 
and here we merely comment that, 
while we have been considering a method 
to construct a metric {\it accurately at an instance}
during the evolution due to the radiation reaction 
under the adiabatic approximation of the orbit, 
we focus on how to construct 
an {\it accurate evolving} metric together with the orbit. 
An accurate derivation of the metric and orbit at an instant 
is a necessary step, but this does not necessarily give us 
the accurate evolving metric and orbit. 
We refer to this as the ``Burke problem''.\footnote{
Professor William Burke was 
a pioneer of the gravitational radiation reaction problem. 
When he was a student at CalTech, he challenged his supervisor, 
Professor Kip Thorne, with this question. 
One can see a record of Burke's challenge 
in front of Thorne's office at CalTech. 
Although he conceded by his famous discovery 
of the so-called Burke-Thorne radiation reaction potential\cite{btf} 
under the post-Newtonian approximation, 
the question of mathematical validity involved 
in the treatment of the radiation reaction problem remains 
a fundamental question in the study of the gravitational radiation reaction.} 

The validity of the second assumption was recently investigated 
in Ref.\cite{phase} for the case of an equal mass binary, 
and in Ref.\cite{adi} for the case of an extreme mass ratio binary. 
The correction to the second assumption is called the ``phasing effect''. 
The evolution equations of the ``constants'' of motion, (\ref{eq:radf}), 
become generally time varying functions for fixed ${\cal E}^a$. 
With the second assumption of the adiabatic approximation, 
we usually take only the part of (\ref{eq:radf}) 
that does not depend explicitly on time. 
Thus, a correction to the second assumption can be obtained 
by calculating the full evolution equation (\ref{eq:radf}), 
which depends explicitly on time. 

This problem was not considered until recently 
in the case of an equal mass binary system, 
because this effect does not appear 
when the relative orbit of the two body system is instantaneously circular 
and because the orbit of an equal mass binary system is circular 
to a very high precision 
in the observationally most interesting case.\cite{cir}
When the orbit is circular, 
the metric and the matter source allow the helical Killing vector, 
$\xi^\alpha=(\partial/\partial t)^\alpha
+\Omega(\partial/\partial \phi)^\alpha$, 
where $\Omega$ is the angular velocity of the orbit 
for a broad range of choices of gauge conditions. 
The $4$-velocity of the orbit is proportional to the helical Killing vector, 
and the radiation reaction acting on the orbit is constant 
along this Killing vector. 
Thus, (\ref{eq:radf}) does not depend explicitly on time 
for a circular orbit. 

This problem has an aspect to which a special attention must be paid. 
Although the time averaged components of (\ref{eq:radf}) 
are gauge invariant, 
the rest of (\ref{eq:radf}) is gauge dependent in general.\footnote{
Strictly speaking, even the time averaged components of (\ref{eq:radf}) 
are gauge dependent. 
However, in a sufficiently large and physically reasonable class 
of gauge conditions, 
we showed in Ref.\cite{adi} that 
the time averaged components of (\ref{eq:radf}) become invariant.} 
Thus, we need the additional argument that 
those gauge dependent components are actually what we can observe 
by calculating gravitational waveforms at the spatial infinity. 

To conclude this section, we summarize the qualitative behavior 
of the orbital evolution due to the radiation reaction 
without making the second assumption 
of the adiabatic approximation.\cite{adi} 
We suppose that the adiabatic metric perturbation in Ref.\cite{adi} 
can be defined in the equal mass binary case 
and that the result of the radiation reaction 
induced by a ``conservative orbit'' can be used to infer 
the evolution of the ``constants'' of motion 
over the radiation reaction time scale. 
The ``constants'' of motion evolve qualitatively as 
\begin{eqnarray}
{\cal E}^a &=& {\cal E}^a_0\left(1+\frac{t}{T_{\hbox{rad}}}+\delta\right) 
\label{eq:Eev} \,, 
\end{eqnarray}
where ${\cal E}^a_0$ are the initial values, 
and $T_{\hbox{rad}}$ is the radiation reaction time 
defined by $T_{\hbox{rad}}={\cal E}/<d{\cal E}/dt>$.\footnote{
Precisely speaking, 
the expressions given in (\ref{eq:Eev}) and (\ref{eq:Pev}) 
are valid only when $t \ll T_{\hbox{rad}}$. 
Terms which behave as $\sim t^2/T_{\hbox{rad}}^2$ 
appear in (\ref{eq:Eev}) when $t \sim T_{\hbox{rad}}$. 
However, it has been argued that as long as we are interested 
in the validity of the adiabatic approximation, 
its qualitative features can be understood sufficiently well 
from (\ref{eq:Eev}) and (\ref{eq:Pev}).\cite{adi}} 
The second term on the RHS of (\ref{eq:Eev}) 
comes from the time averaged components of (\ref{eq:radf}), 
and the third term comes from the rest of (\ref{eq:radf}). 
If we assume that the amplitude of (\ref{eq:radf}) is $1/T_{\hbox{rad}}$ 
and that (\ref{eq:radf}) varies 
on the dynamical time scale of the orbit, $T_{\hbox{dyn}}$ , 
we obtain the relation $\delta \sim T_{\hbox{dyn}}/T_{\hbox{rad}}$. 

This is not the end of the story, however, 
because what we observe is the phase of the gravitational waves, 
which is twice the angular position of the binary. 
Noting that the instantaneous frequency, $\Omega$,  
is a function of the ``constants'' of motion, 
the phase evolves as 
\begin{eqnarray}
\Phi(t) &=& \int dt \Omega({\cal E}^a) 
~=~ \Omega({\cal E}^a_0)\left(t+\frac{t^2}{T_{\hbox{p.rad}}}+\delta_p\right) 
\label{eq:Pev} \,, 
\end{eqnarray}
where $T_{\hbox{p.rad}}=(2\Omega/\partial_a\Omega{\cal E}^a_0)T_{\hbox{rad}}
\sim T_{\hbox{rad}}$. 
Because of the time integration, 
we have $\delta_p \sim T_{\hbox{dyn}}t/T_{\hbox{rad}}$. 

Equation (\ref{eq:Pev}) suggests that $\delta_p$ becomes important 
at the radiation reaction time scale, $t\sim T_{\hbox{p.rad}}$ 
for an accurate prediction of the phase, 
and we need to consider the phasing effect, which is ignored 
when the second assumption of the adiabatic approximation is employed. 
In the case of an equal mass binary, 
it is necessary to calculate the phasing effect 
with an appropriate gauge condition 
to obtain an accurate prediction of the gravitational wave phase 
over the radiation reaction time scale. 
However, as we discuss in Ref.\cite{adi}, 
in the case of an extreme mass ratio binary, 
we may choose a gauge condition 
under which we have $\delta_p \sim T_{\hbox{dyn}}^2/T_{\hbox{rad}}$, 
thus, the phasing effect is not important 
for a reliable prediction of gravitational waveforms.

\section{Gauge problem and first principles calculation} \label{sec:pri}

With the first assumption of the adiabatic approximation, 
we consider first calculating the instantaneous metric 
induced by a conserved orbit as accurately as possible, 
and then deriving the evolution equation for the ``constants'' 
of the form (\ref{eq:radf}) from the instantaneous metric. 
In order to ascertain the validity of the first assumption, 
one has to estimate the extra terms 
coming from the evolving metric 
induced by the orbit deviating from a conservative orbit 
due to the radiation reaction. 
We refer to the derivation of (\ref{eq:radf}) 
together with a consistent derivation of the metric 
as the ``first principles calculation''. 

The most important problem here is properly choosing the gauge condition. 
The gauge condition entirely determines the orbital evolution, 
and it has been shown that one may even choose a gauge condition 
such that the RHS of (\ref{eq:radf}) vanishes 
consistently with an usual metric perturbation scheme.\cite{adi} 
However, such a choice of the gauge condition is not appropriate, 
because the perturbative calculation of the metric becomes 
invalid on a short time scale, 
and consequently, one cannot make any meaningful prediction 
of gravitational waveforms modulated by the radiation reaction.\cite{adi} 
Unfortunately, we have not yet found 
a concise criterion for this gauge condition. 
In order to determine whether or not a gauge condition is appropriate 
to study the radiation reaction problem, 
one must calculate the evolution of the metric 
together with the orbital evolution due to the radiation reaction. 
If the resulting evolution of the metric 
is an accurate approximation for a long time 
(longer than the radiation reaction time scale), 
we can conclude that the gauge condition is appropriate 
to study the radiation reaction problem on this time scale. 

Well known methods for solving the Einstein equation, 
such as the post-Newtonian approximation for an equal mass binary 
and the black hole perturbation for an extreme mass ratio binary, 
are applicable when the binary motion is conservative. 
Therefore, one can derive the instantaneous metric with these methods. 
The evolving metric may be derived 
by applying the adiabatic expansion to this instantaneous metric, 
as we proposed in Ref.\cite{adi}.
Although the original idea 
of the adiabatic metric perturbation given in Ref.\cite{adi} 
seems successful for the study of the radiation reaction 
of an extreme mass ratio binary, 
it was defined only for a linear perturbation, and 
thus, it is not clear whether the idea can be extended 
to the case of an equal mass binary 
for which the non-linear calculation of the metric with a flat background 
is done under the post-Newtonian approximation. 
In addition, the equation of motion 
resulting from the adiabatic metric perturbation 
poses a serious question with regard to its predictability. 

Using the adiabatic approximation, 
we can calculate the instantaneous metric 
induced by a conserved orbit with a certain approximation, 
such as the $2$nd post-Newtonian approximation 
in the case of an equal mass binary 
and the black hole linear perturbation 
in the case of an extreme mass ratio binary. 
The orbital equation for the radiation reaction (\ref{eq:radf}) 
is derived from the instantaneous metric. 
Recall that we restrict our consideration to the case 
that the $4$-velocity is determined 
by the constants of motion, ${\cal E}$. 
The orbital coordinates as functions of the orbital parameter 
can be obtained by further integrating the $4$-velocity, 
and the integral constants for this integration are called 
the ``phase constants'', ${\cal C}$. 
Thus, the conserved orbit is determined by ${\cal E}$ and ${\cal C}$, 
and the instantaneous metric is derived 
as a function of ${\cal E}$ and ${\cal C}$. 
As a result, the RHS of (\ref{eq:radf}) 
becomes a function of ${\cal E}$ and ${\cal C}$ 
under the adiabatic approximation in general. 
Applying an appropriate isometry transformation, 
one can make the phase constants vanish 
in the case of an equal mass binary. 
We showed that this is true for the extreme mass ratio binary case 
in Ref.\cite{adi}. 
Therefore, under the adiabatic approximation of the orbit, 
the RHS of (\ref{eq:radf}) is a function of only ${\cal E}$, 
and one can derive the evolution of ${\cal E}$ 
as a function of the orbital parameter 
in a closed form using (\ref{eq:radf}). 

Now, let us consider a correction to the adiabatic approximation. 
We consider the orbital evolution due to the radiation reaction; 
that is, ${\cal E}$ and ${\cal C}$ are no longer constant. 
Then, it becomes necessary to consider 
the explicit dependence on ${\cal C}$ of the metric, 
because the isometry transformation cannot eliminate ${\cal C}$ 
during the evolution. 
Furthermore, the dependence on ${\cal C}$ is more important 
than that on ${\cal E}$, 
because ${\cal C}$ evolves more rapidly than ${\cal E}$.\cite{adi} 
In Ref.\cite{adi}, preliminary discussion on this effect is given 
for the case of an extreme mass ratio binary. 
We define the adiabatic linear metric perturbation 
as that consisting of the replacement 
of the constants ${\cal E}$ and ${\cal C}$ 
by the evolving ${\cal E}$ and ${\cal C}$ 
of the instantaneous metric. 
Through this operation, 
the evolution equation (\ref{eq:radf}) becomes that 
which we use under the adiabatic approximation of the orbit, 
to leading order in the perturbation expansion, 
and it does not depend on ${\cal C}$. 
Next we consider the equation for the correction 
to the adiabatic linear metric perturbation. 
Although we could not obtain 
a consistent expansion scheme in Ref.\cite{adi}, 
the source term for the correction 
to the adiabatic linear metric perturbation 
includes terms proportional to the first and second derivatives 
of ${\cal E}$ and ${\cal C}$. 
If we consider the linear correction 
of the adiabatic linear metric perturbation, 
it may be reasonable that the correction metric 
includes terms proportional to the first and the second derivatives 
of ${\cal E}$ and ${\cal C}$. 
In this case, instead of (\ref{eq:radf}), 
the orbital equation becomes 
\begin{eqnarray}
\frac{d{\cal E}^a}{dt} = \tilde {\cal F}^a
({\cal E},\dot{\cal E},\ddot{\cal E},\dot{\cal C},\ddot{\cal C}) 
\,, \label{eq:radf1} 
\end{eqnarray}
and one cannot solve the orbital evolution for ${\cal E}$ 
in a closed form. 

One can eliminate the dependences on 
$\dot{\cal E}$, $\ddot{\cal E}$, $\dot{\cal C}$ and $\ddot{\cal C}$ 
using the orbital equation 
derived with the adiabatic linear metric perturbation. 
In Ref.\cite{adi}, we derived the qualitative behavior 
of the derivatives of ${\cal E}$ and ${\cal C}$ 
employing the adiabatic linear metric perturbation as 
\begin{eqnarray}
&& \dot{\cal E} ~\sim~ {\cal E}\frac{1}{T_{\hbox{rad}}} \,, \quad 
\ddot{\cal E} ~\sim~ {\cal E}\frac{1}{T_{\hbox{rad}}^2} \,, 
\nonumber \\ 
&& \dot{\cal C} ~\sim~ \frac{\partial\Omega}{\partial{\cal E}}
\left({\cal E}-{\cal E}_0\right) \,, \quad 
\ddot{\cal C} ~\sim~ 
\frac{\partial\Omega}{\partial{\cal E}}{\cal E}\frac{1}{T_{\hbox{rad}}} 
\nonumber \,, 
\end{eqnarray}
where $\Omega$ is the frequency associated with the phase constants, 
which is a function of ${\cal E}$, 
and ${\cal E}_0$ is the initial value of ${\cal E}$ 
when we switch on the radiation reaction. 
Because the behavior of $\dot{\cal C}$ involves ${\cal E}_0$, 
we have two serious problems, described below. \par 
\noindent {\it 1) Adiabaticity:} The orbital prediction obtained this equation 
depends on the choice of the initial value, ${\cal E}_0$. \par 
\noindent {\it 2) The Burke problem:} Because $\dot{\cal C}$ grows in time, 
the correction to the adiabatic linear metric perturbation 
may also grow in time. 
As a result, the correction would eventually become dominant 
over the adiabatic metric. \par
\noindent The first problem suggests that 
we might not be able to predict the gravitational waveform, 
because the result depends on the choice of the initial time. 
This would imply that our basic hypothesis of 
the adiabatic evolution of the orbit might be wrong. 
The second problem suggests that 
the adiabatic linear metric perturbation may not 
be a good approximation of the metric at large times. 

These problems are unexpected with our current understanding 
of the gravitational radiation reaction problem. 
Actually, it is widely accepted by the physics community 
that the orbital evolution calculated under the adiabatic approximation 
was confirmed by a Nobel Prize winning observation 
of the Hulse-Taylor binary pulsar.\cite{HTb} 

We believe that the cause of these problems 
is an inappropriate choice of the gauge condition. 
As we stressed in Ref.\cite{adi}, 
the calculation could become highly nonperturbative 
with an inappropriate choice of the gauge condition 
when we consider the evolution of the metric and the orbit 
due to the radiation reaction. 
For this reason, we cannot predict the gravitational waveform 
for a sufficiently long time. 
This is the Burke problem. 
If the Burke problem arises, 
the prediction is valid for only a short time, 
and the initial time can be chosen only within this short time span. 
This is the adiabaticity problem. 
We note that the source term 
for the correction to the adiabatic linear metric perturbation 
depends on the gauge condition. 
We hypothesize that an appropriate gauge condition is the condition 
that we can eliminate the $\dot{\cal C}$-dependence of the source term 
and we may be able to avoid these problems. 

In the following two sections, we propose 
a consistent perturbation scheme to construct 
the adiabatic metric in a nonlinear manner 
by extending the idea proposed in Ref.\cite{adi} 
This scheme consists of two different perturbation methods 
with the same small expansion parameter $\lambda$. 
In \S \ref{sec:bur}, we show that, 
under a specific gauge condition, 
one can actually avoid the Burke problem 
for the leading-order correction 
to the adiabatic linear metric perturbation.

\section{Renormalized metric perturbation} \label{sec:rmp}

We consider the Einstein equation, 
employing the geometrical units, in which$G=c=1$. 
\begin{eqnarray}
G^{\alpha\beta} &=& 8\pi T^{\alpha\beta} \,. 
\end{eqnarray}
In this section, we discuss the formal application 
of the renormalized perturbation method to a metric perturbation, 
and we do not specify the form of the stress-energy tensor. 
We consider a general vacuum metric $g^{(bk)\alpha\beta}$ 
as the background for the perturbative expansion, 
so that one can use this scheme 
for both the equal mass binary case and the extreme mass ratio binary case. 
Below, the covariant derivative is taken with respect to the background. 

The expansion parameter $\lambda$ is taken to be 
the amplitude of the stress-energy tensor, 
$T^{\alpha\beta} \sim O(\lambda)$. 
Because the metric perturbation induced by this stress-energy tensor 
creates a radiation reaction that influences the evolution of the matter, 
we can assume $T^{\alpha\beta}{}_{;\beta} \sim O(\lambda^2)$. 

We expand the metric in $\lambda$ as 
\begin{eqnarray}
g^{\alpha\beta} &=& g^{(bk)\alpha\beta}
+\lambda h^{(1)\alpha\beta}+\lambda^2 h^{(2)\alpha\beta}
+\lambda^3 h^{(3)\alpha\beta}+\cdots 
\,. \label{eq:r_mp} 
\end{eqnarray}
The Einstein tensor can be schematically expanded as 
\begin{eqnarray}
G^{\alpha\beta} &=& \lambda \left[G^{\alpha\beta}\right]^{(1)}(h^{(1)})
\nonumber \\ 
&&+\lambda^2 \left\{\left[G^{\alpha\beta}\right]^{(1)}(h^{(2)})
+\left[G^{\alpha\beta}\right]^{(2)}(h^{(1)},h^{(1)})\right\}
\nonumber \\ 
&&+\lambda^3 \left\{\left[G^{\alpha\beta}\right]^{(1)}(h^{(3)})
+2\left[G^{\alpha\beta}\right]^{(2)}(h^{(1)},h^{(2)})
+\left[G^{\alpha\beta}\right]^{(3)}(h^{(1)},h^{(1)},h^{(1)})\right\}
\nonumber \\ 
&&+\cdots \,, 
\end{eqnarray}
where $\left[G^{\alpha\beta}\right]^{(1)}(h)$ is linear in $h^{\alpha\beta}$, 
$\left[G^{\alpha\beta}\right]^{(2)}(h^{(1)},h^{(2)})$ is bi-linear 
in both $h^{(1)\alpha\beta}$ and $h^{(2)\alpha\beta}$, and so on. 
We recall that there is an algebraic relation 
$\left[G^{\alpha\beta}\right]^{(1)}_{;\beta}(h)=0$ 
for any $h^{\alpha\beta}$. 

To obtain a consistent perturbation expansion, 
we could consider the expansion of the stress-energy tensor, 
$T^{\alpha\beta}$, in $\lambda$. 
If we do this, the stress-energy tensor 
is conserved in the background to leading order due to this consistency, 
and the dissipative dynamics of the matter due to the radiation reaction 
can only be taken into account at higher order. 
Because the dissipative effect accumulates in time, 
the higher-order terms eventually become dominant over the leading term, 
and the expansion of the stress-energy tensor soon becomes invalid. 
This implies that the Burke problem is inevitable, 
and we cannot make a physically meaningful prediction 
of the system's evolution due to the radiation reaction.\cite{adi} 
For this reason, we do not expand the stress-energy tensor 
and propose a new perturbation scheme 
with which the dissipative dynamics of the stress-energy tensor 
can be renormalized in the leading order of the metric perturbation 
in a consistent manner. 
For this purpose, we use a gauge fixing term 
as a counterterm for the renormalized perturbation. 
This idea is similar 
to that of the so-called fast motion approximation.\cite{fm} 
However, the key difference is that 
in the present case, we have order-by-order equations 
for the metric perturbations at each order 
in the adiabatic expansion we propose in the next section. 

It is important to note that 
the stress-energy tensor does not satisfy 
the conservation equation in the background, 
$T^{\alpha\beta}{}_{;\beta}\not =0$, 
due to the radiation reaction caused by the metric perturbation. 
Therefore, it is inconsistent to simply equate 
$\lambda\left[G^{\alpha\beta}\right]^{(1)}(h^{(1)})$ 
with $T^{\alpha\beta}$ 
in order to define $h^{(1)\alpha\beta}$. 
We also note that, for a given divergence-free tensor 
$f^{\alpha\beta}\,(f^{\alpha\beta}{}_{;\beta}=0)$, 
the linear Einstein equation 
$\left[G^{\alpha\beta}\right]^{(1)}(h)=f^{\alpha\beta}$ 
with an appropriate boundary condition 
does not have a unique solution, because of the gauge freedom. 
This is equivalent to the fact that the linearized Einstein operator 
$\left[G^{\alpha\beta}\right]^{(1)}_{\gamma\delta}$ 
does not have an inverse (the Green function) 
until we add a gauge fixing operator. 
We formally define the gauge fixing operator 
$\Lambda^{\alpha\beta}{}_{\gamma\delta}$, 
which consists of differential operators that are at most second order. 
We have a tensor Green function 
$g^{\alpha\beta}{}_{\gamma'\delta'}(x,x')$ that satisfies 
\begin{eqnarray}
&& \left\{\left[G^{\alpha\beta}\right]^{(1)}_{\gamma\delta}
+\left[\Lambda^{\alpha\beta}\right]_{\gamma\delta}\right\}
\cdot g^{\gamma\delta}{}_{\eta'\zeta'}(x,x') 
\nonumber \\ && \qquad 
~=~ \frac{1}{2}(\delta^\alpha{}_{\eta'}\delta^\beta{}_{\zeta'}
+\delta^\alpha{}_{\zeta'}\delta^\beta{}_{\eta'})
\frac{\delta^{(4)}(x-x')}{\sqrt{-g^{(bk)}}}
\,, \label{eq:g_ein} 
\end{eqnarray}
where $\delta^{(4)}(x-x')$ is the $4$-dimensional Dirac delta function 
and $g^{(bk)}$ is the matrix determinant of $g^{(bk)}_{\alpha\beta}$. 
We choose the gauge fixing term such that 
the linearized Einstein equation is 
no longer divergence free in an algebraic manner 
and, as a result, it allows a general symmetric tensor as a source. 
We remark that the Green function defined in this manner includes 
propagation of not only the gravitational mode 
but also the vector and the scalar modes. 

We define the linear metric perturbation 
$\lambda h^{(1)\alpha\beta}$ of (\ref{eq:r_mp}) as 
\begin{eqnarray}
\lambda h^{(1)\alpha\beta} &=& 
\int d^4x' \sqrt{-g^{(bk)}}g^{\alpha\beta}{}_{\gamma'\delta'}(x,x') 
8\pi T^{c'd'}(x') \,. \label{eq:r_mp1} 
\end{eqnarray}
By definition, the linear metric perturbation satisfies 
\begin{eqnarray}
\left[G^{\alpha\beta}\right]^{(1)}(\lambda h^{(1)}) &=& 
8\pi T^{\alpha\beta} -\Lambda^{\alpha\beta}(\lambda h^{(1)}) 
\,. \label{eq:r_ein1} 
\end{eqnarray}
We recall that $T^{\alpha\beta}{}_{;\beta} \sim O(\lambda^2)$ 
and we require that $\lambda h^{(1)\alpha\beta}$ be 
a solution of the linear Einstein equation to $O(\lambda)$. 
Then, we have the condition 
\begin{eqnarray}
\Lambda^{\alpha\beta}(\lambda h^{(1)}) &\sim& O(\lambda^2) 
\,. \label{eq:r_rq1} 
\end{eqnarray}
Taking the divergence of (\ref{eq:r_ein1}), we have 
\begin{eqnarray}
T^{\alpha\beta}{}_{;\beta} &=& 
\frac{1}{8\pi}\Lambda^{\alpha\beta}{}_{;\beta}(\lambda h^{(1)}) 
\,\sim\, O(\lambda^2) 
\label{eq:r_eom1} \,, 
\end{eqnarray}
which gives the equation of motion to an accuracy of $O(\lambda)$ 
as a consistency condition. 

The gauge fixing term of (\ref{eq:r_ein1}) 
must be taken into account as a source 
for the second-order metric perturbation, 
together with the quadratic contribution 
of the linear metric perturbation. 
Using the tensor Green function, 
we define the second-order metric perturbation as 
\begin{eqnarray}
\lambda^2 h^{(2)\alpha\beta} &\equiv& 
\int d^4x' g^{\alpha\beta}{}_{\gamma'\delta'}(x,x') 
\nonumber \\ && \quad 
\times \left(\Lambda^{\gamma'\delta'}(\lambda h^{(1)})(x')
-\left[G^{\gamma'\delta'}\right]^{(2)}(\lambda h^{(1)},\lambda h^{(1)})(x')
\right) \label{eq:r_mp2} \,. 
\end{eqnarray}
By definition, the second-order metric perturbation satisfies 
\begin{eqnarray}
&& \left[G^{\alpha\beta}\right]^{(1)}(\lambda^2 h^{(2)}) 
+\left[G^{\alpha\beta}\right]^{(2)}(\lambda h^{(1)},\lambda h^{(1)}) 
\nonumber \\ && \qquad 
~=~ \Lambda^{\alpha\beta}(\lambda h^{(1)}) 
-\Lambda^{\alpha\beta}(\lambda^2 h^{(2)}) 
\label{eq:r_ein2} \,. 
\end{eqnarray}
As (\ref{eq:r_rq1}), in order to make $\lambda^2 h^{(2)\alpha\beta}$ 
an approximate second-order metric perturbation to $O(\lambda^2)$, 
we impose the condition 
\begin{eqnarray}
\Lambda^{\alpha\beta}(\lambda^2 h^{(2)}) &\sim& O(\lambda^3) 
\,. \label{eq:r_rq2}
\end{eqnarray}
Taking the divergence of (\ref{eq:r_ein2}), we have 
\begin{eqnarray}
T^{\alpha\beta}{}_{;\beta} &=& 
\frac{1}{8\pi}\Lambda^{\alpha\beta}{}_{;\beta}(\lambda h^{(1)}) 
\nonumber \\ 
&=& \frac{1}{8\pi}
\left[G^{\alpha\beta}\right]^{(2)}(\lambda h^{(1)},\lambda h^{(1)})_{;\beta} 
+O(\lambda^3) 
\label{eq:r_eom2} \,. 
\end{eqnarray}
This shows that the equation of motion to $O(\lambda^2)$ 
can be derived with the linear metric perturbation 
$\lambda h^{(1)\alpha\beta}$. 

In a similar manner, we define the $n$-th order metric perturbations as 
\begin{eqnarray}
\lambda^n h^{(n)\alpha\beta} &=& 
\int d^4x' g^{\alpha\beta}{}_{\gamma'\delta'}(x,x') 
\nonumber \\ && \qquad 
\times \biggl(
\Lambda^{\gamma'\delta'}(\lambda h^{(n-1)})(x')
-2\left[G^{\gamma'\delta'}\right]^{(2)}
(\lambda^{n-1} h^{(n-1)},\lambda h^{(1)})(x') 
\nonumber \\ && \qquad \quad 
-\cdots
-\left[G^{\gamma'\delta'}\right]^{(n)}(\lambda h^{(1)},\cdots)(x')
\biggr) \label{eq:r_mpx} \,. 
\end{eqnarray}
This satisfies the equation 
\begin{eqnarray}
&& \left[G^{\alpha\beta}\right]^{(1)}(\lambda^n h^{(n)}) 
+\cdots 
+\left[G^{\alpha\beta}\right]^{(n)}(\lambda h^{(1)},\cdots)
\nonumber \\ && \qquad 
~=~ \Lambda^{\alpha\beta}(\lambda^{n-1} h^{(n-1)}) 
-\Lambda^{\alpha\beta}(\lambda^n h^{(n)}) 
\label{eq:r_einx} \,. 
\end{eqnarray}
By imposing the condition 
\begin{eqnarray}
\Lambda^{\alpha\beta}(\lambda^n h^{(n)}) &\sim& O(\lambda^{n+1}) 
\,, \label{eq:r_schx}
\end{eqnarray}
we obtain the equation of motion to $O(\lambda^n)$ as 
\begin{eqnarray}
T^{\alpha\beta}{}_{;\beta} &=& \frac{1}{8\pi}\biggl(
\left[G^{\alpha\beta}\right]^{(2)}
(\lambda h^{(1)},\lambda h^{(1)})_{;\beta} 
\nonumber \\ && \qquad 
+2\left[G^{\alpha\beta}\right]^{(2)}
(\lambda^2 h^{(2)},\lambda h^{(1)})_{;\beta} 
+\left[G^{ab}\right]^{(3)}
(\lambda h^{(1)},\lambda h^{(1)},\lambda h^{(1)})_{;b} 
\nonumber \\ && \qquad 
+\cdots 
\nonumber \\ && \qquad 
+2\left[G^{ab}\right]^{(2)}(\lambda^{n-1} h^{(n-1)},\lambda h^{(1)})_{;b} 
+\cdots 
+\left[G^{ab}\right]^{(n)}(\lambda h^{(1)},\lambda h^{(1)},\cdots)_{;b} 
\nonumber \\ && \qquad 
+O(\lambda^{n+1}) \biggr)
\label{eq:r_eomx} \,. 
\end{eqnarray}
The equation of motion to $O(\lambda^n)$ 
can be derived by calculating the metric perturbations, 
$\lambda h^{(k)\alpha\beta} \,(k=1,\cdots,n-1)$. 

We note that the gauge condition 
is obscure in this perturbation scheme. 
This is because we do not expand the stress-energy tensor in $\lambda$, 
and as a result, the stress-energy tensor does not satisfy 
the conservation law in the background metric. 
We do not equate 
$\lambda\left[G^{\alpha\beta}\right]^{(1)}(h^{(1)})$ 
and $T^{\alpha\beta}$, as in the usual metric perturbation scheme, 
and, as we see from (\ref{eq:r_ein1}), 
we incorporate the difference of these terms in the gauge fixing term. 
Therefore, the gauge fixing term does not vanish 
in this perturbation scheme. 

We also note that the validity of this perturbation scheme 
depends on the gauge fixing term. 
In order to see whether or not 
we can avoid the Burke problem considered in \S \ref{sec:pri}, 
we must explicitly check the behavior of the gauge fixing term. 
In \S \ref{sec:bur}, we consider this problem 
with a specific gauge fixing term.

\section{Adiabatic expansion} \label{sec:amp}

Following the formal discussion of the metric perturbation scheme 
given in the previous section, 
we define the metric perturbation 
which includes the radiation reaction effect to leading order 
in a consistent manner 
as (\ref{eq:r_mp1}), (\ref{eq:r_mp2}) and (\ref{eq:r_mpx}). 
We may still have a technical difficulty 
arising in an explicit calculation of these metric perturbations. 
Because the source term for $h^{(1)\alpha\beta}$ cannot be derived 
until we derive the equation of motion to a necessary order, 
the metric perturbation scheme presented in the previous section 
is not systematic in the sense that 
one cannot derive a metric perturbation of a given order 
from lower order metric perturbations. 
A similar difficulty appears 
in the so-called fast motion approximation.\cite{fm} 
In order to avoid this technical problem, 
we further use another perturbation scheme proposed in Ref.\cite{adi} 
which uses the same small parameter, $\lambda$. 
For simplicity, we only discuss the derivation of 
$h^{(1)\alpha\beta}$ by integrating (\ref{eq:r_ein1}). 
We consider the application of the adiabatic expansion 
proposed in Ref.\cite{adi} 
with $h^{(1)\alpha\beta}$ of the form 
\begin{eqnarray}
\lambda h^{(1)\alpha\beta} &=& \lambda h^{(1,0)\alpha\beta}
+\lambda^2 h^{(1,1)\alpha\beta}+\cdots \,. \label{eq:a_mp} 
\end{eqnarray}
We suppose that we are able to calculate 
the higher-order metric perturbations 
(\ref{eq:r_mp2}) and (\ref{eq:r_mpx}) 
in the similar manner through the expansion 
\begin{eqnarray}
\lambda^n h^{(n)\alpha\beta} &=& \lambda^n h^{(n,0)\alpha\beta}
+\lambda^{n+1} h^{(n,1)\alpha\beta}+\cdots \,.
\end{eqnarray}

We consider the stress-energy tensor of point particles 
and restrict ourselves to the case in which 
the equation of motion (\ref{eq:r_eomx}) is integrable, 
with constants of motion ${\cal E}^a$ 
in the absence of a gravitational radiation reaction, 
as in the case of a binary system of non-spinning particles. 
As discussed in \S \ref{sec:pri}, 
we also need the phase constants ${\cal C}^b$ 
in order to specify the orbital coordinates, 
and the conserved stress-energy tensor 
is described as a function of ${\cal E}^a$ and ${\cal C}^b$. 
We suppose that there is a method to calculate (\ref{eq:r_mp1}) 
for a conserved stress-energy tensor. 
Using this method, we have the ``ordinary'' linear metric perturbation 
$\tilde h^{(1)\alpha\beta}(x;{\cal E}^a,{\cal C}^b)$, 
which is the solution of the usual linearized Einstein equation 
$\lambda\left[G^{\alpha\beta}\right]^{(1)}(\tilde h^{(1)})
=T^{\alpha\beta}(x;{\cal E}^a,{\cal C}^b)$. 
(Because we do not consider the radiation reaction at this stage, 
the stress-energy tensor on the RHS is conserved in the background.) 

Now we consider the effect of the radiation reaction. 
Assuming the adiabatic evolution of the orbits, 
we consider the ``constants'' as functions of the local time $t$ 
which evolve due to the radiation reaction. 
Assuming the small parameter $\lambda$ used in \S \ref{sec:rmp}, 
we have $\frac{d}{dt}{\cal E}^a \sim 
\frac{d}{dt}{\cal C}^b \sim O(\lambda)$. 
In Ref.\cite{adi}, we extend 
this picture of the adiabatic evolution of the orbit 
to the evolution of the linear metric perturbation. 
We foliate the spacetime into smooth spacelike hypersurfaces, 
and the foliation function, $f(x)$, is taken 
such that it coincides the local time $t$ around the particles. 
The adiabatic linear metric perturbation is defined 
by using the ``ordinary'' linear metric perturbation 
of the constants, ${\cal E}^a(f)$ and ${\cal C}^b(f)$, 
on the foliation surface, $f(x)=f$. 
Because $\lim_{\lambda\to 0}(T^{\alpha\beta}_{;\beta}/\lambda)\to 0$, 
the adiabatic metric perturbation gives (\ref{eq:r_mp1}) to $O(\lambda)$. 
Therefore, we take the adiabatic linear metric perturbation 
as the leading term in the adiabatic expansion of (\ref{eq:a_mp}): 
\begin{eqnarray}
h^{(1,0)\alpha\beta} &=& 
\tilde h^{(1)\alpha\beta}(x;{\cal E}^a(f(x)),{\cal C}^b(f(x))) 
\,. \label{eq:a_mp0} 
\end{eqnarray}
This yields the orbital equation (\ref{eq:r_eom2}) to $O(\lambda)$ as 
\begin{eqnarray}
\frac{d}{dt}{\cal E}^a ~=~ 
\lambda\left[\frac{d}{dt}{\cal E}^a\right]^{(1)}+O(\lambda^2) 
\,, \quad 
\frac{d}{dt}{\cal C}^b ~=~ 
\lambda\left[\frac{d}{dt}{\cal C}^b\right]^{(1)}+O(\lambda^2) 
\,. \label{eq:a_eom1}
\end{eqnarray}

The equation for the correction to (\ref{eq:a_mp0}) 
can be obtained by inserting (\ref{eq:a_mp}) into (\ref{eq:r_ein1}). 
Because (\ref{eq:r_ein1}) is linear in $h^{(1)\alpha\beta}$, 
we have 
\begin{eqnarray}
&& \sum_{n=1,2,\cdots}\left\{
\left[G^{\alpha\beta}\right]^{(1)}(\lambda^n h^{(1,n)}) 
+\Lambda^{\alpha\beta}(\lambda^n h^{(1,n)})\right\}
\nonumber \\ && \qquad 
~=~ -\left[G^{\alpha\beta}\right]^{(1)}(\lambda h^{(1,0)}) 
-\Lambda^{\alpha\beta}(\lambda h^{(1,0)})
+T^{\alpha\beta}(x;{\cal E}^a(t),{\cal C}^b(t)) 
\,. \label{eq:a_einx}
\end{eqnarray}
We recall that the linearized Einstein operator 
and the gauge fixing operator consist 
of differential operators that are at most second order. 
Therefore, the RHS of (\ref{eq:a_einx}) becomes 
\begin{eqnarray}
&& \lambda\left(S^{\alpha\beta}_{aa'}
\frac{d{\cal E}^a}{dt}\frac{d{\cal E}^{a'}}{dt}
+S^{\alpha\beta}_{ab}
\frac{d{\cal E}^a}{dt}\frac{d{\cal C}^b}{dt}
+S^{\alpha\beta}_{bb'}
\frac{d{\cal C}^a}{dt}\frac{d{\cal C}^{b'}}{dt}
\right)\nabla_\alpha f(x)\nabla_\beta f(x)
\nonumber \\ && 
+\lambda\left(S^{\alpha\beta}_a\frac{d^2{\cal E}^a}{dt^2}
+S^{\alpha\beta}_b\frac{d^2{\cal C}^b}{dt^2}\right)
\nabla_\alpha\nabla_\beta f(x)
\nonumber \\ && 
+\lambda\left(S^\alpha_a\frac{d{\cal E}^a}{dt}
+S^\alpha_b\frac{d{\cal C}^b}{dt}\right)
\nabla_\alpha f(x) \,, 
\end{eqnarray}
where ${d^n{\cal E}^a}/{dt^n}$ and ${d^n{\cal C}^b}/{dt^n}$ 
are evaluated at $t=f(x)$. 
Using (\ref{eq:a_eom1}), 
we obtain the following equation for $h^{(1,1)\alpha\beta}$ 
by taking the terms of $O(\lambda^2)$ of (\ref{eq:a_einx}): 
\begin{eqnarray}
&& \left[G^{\alpha\beta}\right]^{(1)}(\lambda^2 h^{(1,1)}) 
+\Lambda^{\alpha\beta}(\lambda^2 h^{(1,1)})
\nonumber \\ && \qquad 
~=~ \lambda^2\left(S^{\alpha\beta}_a
\frac{d}{dt}\left[\frac{d}{dt}{\cal E}^a\right]^{(1)}
+S^{\alpha\beta}_b
\frac{d}{dt}\left[\frac{d}{dt}{\cal C}^b\right]^{(1)}\right)
\nabla_\alpha\nabla_\beta f(x)
\nonumber \\ && \qquad \quad 
+\left(S^\alpha_a
\left[\frac{d}{dt}{\cal E}^a\right]^{(1)}
+S^\alpha_b
\left[\frac{d}{dt}{\cal C}^b\right]^{(1)}\right)
\nabla_\alpha f(x) 
\,. \label{eq:a_ein1}
\end{eqnarray}

Thus, we have the equation for $h^{(1,1)\alpha\beta}$, 
where source term is derived using 
the metric perturbation $h^{(1,0)\alpha\beta}$ 
and the orbital equation of motion, 
$\left[\frac{d}{dt}{\cal E}^a\right]^{(1)}$ and 
$\left[\frac{d}{dt}{\cal C}^b\right]^{(1)}$. 
However, (\ref{eq:a_ein1}) cannot be integrated, 
because $h^{(1,0)\alpha\beta}$ depends on 
${\cal E}^a(t)$ and ${\cal C}^b(t)$. 
We consider repeating the adiabatic calculation 
used for $h^{(1,0)\alpha\beta}$ 
in order to integrate (\ref{eq:a_ein1}). 
In other words, instead of solving (\ref{eq:a_ein1}) exactly, 
we integrate the equation, 
regarding ${\cal E}^a$ and ${\cal C}^b$ on the RHS as constants. 
Then, we replace ${\cal E}^a$ and ${\cal C}^b$ 
by ${\cal E}^a(f)$ and ${\cal C}^b(f)$. 
By this method, we can solve (\ref{eq:a_ein1}) 
to an accuracy of $O(\lambda^2)$, 
and we take into account the extra term of $O(\lambda^3)$ 
coming from this adiabatic calculation of $h^{(1,1)\alpha\beta}$ 
as an additional source term for $h^{(1,2)\alpha\beta}$. 

The equation for $h^{(1,2)\alpha\beta}$ can be derived 
from the metric perturbations 
$h^{(1,0)\alpha\beta}$ and $h^{(1,1)\alpha\beta}$, 
and the orbital equations 
$\left[\frac{d}{dt}{\cal E}^a\right]^{(1)}$, 
$\left[\frac{d}{dt}{\cal E}^a\right]^{(2)}$, 
$\left[\frac{d}{dt}{\cal C}^b\right]^{(1)}$ and 
$\left[\frac{d}{dt}{\cal C}^b\right]^{(2)}$. 
In order to derive $\left[\frac{d}{dt}{\cal E}^a\right]^{(2)}$ 
and $\left[\frac{d}{dt}{\cal C}^b\right]^{(2)}$, 
$h^{(1,1)\alpha\beta}$ is insufficient, 
and it is necessary to calculate $h^{(2,0)\alpha\beta}$, 
as we see from (\ref{eq:r_eomx}) with $n=2$. 
The equation for $h^{(2,0)\alpha\beta}$ can be derived 
from $h^{(1,0)\alpha\beta}$. 
Thus, $h^{(1,2)\alpha\beta}$ can be derived 
from lower-order metric perturbations. 

Repeating this procedure, 
we find that the equation for $h^{(n,m)\alpha\beta}$ 
can be derived from the metric perturbations 
$h^{(n',m')\alpha\beta}~(n'+m'<n+m)$ 
and the equations of motion 
$\left[\frac{d}{dt}{\cal E}^a\right]^{(n')}$ 
and $\left[\frac{d}{dt}{\cal C}^b\right]^{(n')}~(n'<n+m)$. 
The equations of motion 
$\left[\frac{d}{dt}{\cal E}^a\right]^{(n)}$ 
and $\left[\frac{d}{dt}{\cal C}^b\right]^{(n)}$ 
are derived from the metric perturbations 
$h^{(n',m')\alpha\beta}~(n'+m'\leq n)$. 
Thus, with this metric perturbation formalism, 
we can systematically calculate the metric perturbations order by order.

\section{Gauge condition and the Burke problem} \label{sec:bur}

To this point, we have discussed the new metric perturbation scheme 
in \S \ref{sec:rmp} and \S \ref{sec:amp} 
without specifying the gauge condition, 
and there is a possibility that the Burke problem exists 
with an inappropriate gauge condition. 
In this case, although the expansion of the metric 
in the small parameter $\lambda$ is initially consistent, 
it would become invalid 
as the system evolves for a sufficiently long time 
after we switch on the radiation reaction. 
In this section, we choose a specific gauge fixing term 
in the new metric perturbation scheme 
and show that we can avoid the Burke problem 
with this gauge condition up to second order in the metric perturbation. 

We define the gauge fixing term as 
\begin{eqnarray}
\Lambda^{\alpha\beta}{}_{\gamma\delta}
&=& \frac{1}{2}\left(
\nabla^\alpha \nabla_\delta g^{(bk)\beta}_\gamma
+\nabla^\beta \nabla_\delta g^{(bk)\alpha}_\gamma 
-g^{(bk)\alpha\beta}\nabla_\gamma \nabla_\delta\right)
\,. \label{eq:l_gau}
\end{eqnarray}
Then, the renormalized perturbation equations 
for $h^{(1)\alpha\beta}$ and $h^{(2)\alpha\beta}$ become 
\begin{eqnarray}
&& \frac{\lambda}{2}\left(
h^{(1)\alpha\beta;\gamma}{}_{;\gamma}
+2R^\alpha{}_\gamma{}^\beta{}_\delta h^{(1)\gamma\delta}
\right) ~=~ 8\pi T^{\alpha\beta} 
\,, \label{eq:l_ein1} \\ 
&& \frac{\lambda^2}{2}\left(
h^{(2)\alpha\beta;\gamma}{}_{;\gamma}
+2R^\alpha{}_\gamma{}^\beta{}_\delta h^{(2)\gamma\delta}
\right) 
\nonumber \\ && \qquad 
~=~ -\left[G^{\alpha\beta}\right]^{(2)}
(\lambda h^{(1)},\lambda h^{(1)})
+\Lambda^{\alpha\beta}(\lambda h^{(1)}) 
\,, \label{eq:l_ein2} 
\end{eqnarray}
where the Riemann tensor, $R^\alpha{}_\gamma{}^\beta{}_\delta$, 
is defined for the background metric. 
Because (\ref{eq:l_ein1}) does not necessarily require 
the divergence-free condition for $T^{\alpha\beta}$, 
the Green function of (\ref{eq:l_ein1}) includes 
the propagation of not only the gravitational mode 
but also the vector mode and the scalar mode 
as discussed in \S \ref{sec:rmp}. 

We first treat $h^{(1,1)\alpha\beta}$ 
by carrying out the adiabatic expansion of (\ref{eq:l_ein1}). 
Let us denote the Green function 
with an appropriate boundary condition 
by $g^{\alpha\beta}{}_{\gamma'\delta'}(x,x')$. 
We recall that the stress-energy tensor is 
a function of the ``constants'' evolving due to the radiation reaction, 
$T^{\alpha\beta}=T^{\alpha\beta}(x;{\cal E}(t),{\cal C}(t))$. 
In the adiabatic expansion presented in \S \ref{sec:amp}, 
$h^{(1,0)\alpha\beta}$ can be defined in terms of the Green function as 
\begin{eqnarray}
\lambda h^{(1,0)\alpha\beta}(x) &=& 
8\pi\left[\int dx'\sqrt{-g^{(bk)}}
g^{\alpha\beta}{}_{\gamma'\delta'}(x,x')
T^{\gamma'\delta'}(x';{\cal E}(f),{\cal C}(f))
\right]_{f=f(x)}
\,. \label{eq:l_mp10} 
\end{eqnarray}
Then, the equation for $\lambda^2 h^{(1,1)\alpha\beta}(x)$ becomes 
\begin{eqnarray}
&& \frac{\lambda^2}{2}\left(
h^{(1,1)\alpha\beta;\gamma}{}_{;\gamma}
+2R^\alpha{}_\gamma{}^\beta{}_\delta h^{(1,1)\gamma\delta}
\right) 
\nonumber \\ && \qquad 
~=~ 8\pi T^{\alpha\beta} 
-\frac{\lambda}{2}\left(
h^{(1,0)\alpha\beta;\gamma}{}_{;\gamma}
+2R^\alpha{}_\gamma{}^\beta{}_\delta h^{(1,0)\gamma\delta}
\right)+O(\lambda^3) 
\,. \label{eq:l_ein11} 
\end{eqnarray}
Using (\ref{eq:l_mp10}), the RHS of (\ref{eq:l_ein11}) becomes 
\begin{eqnarray}
&& -8\pi\left[\int dx'\sqrt{-g^{(bk)}}
g^{\alpha\beta}{}_{\gamma'\delta'}(x,x')
\frac{d}{df}T^{\gamma'\delta'}(x';{\cal E}(f),{\cal C}(f))
\right]_{f=f(x)} f^{;\epsilon}{}_{;\epsilon}(x)
\nonumber \\ 
&& \quad -8\pi\left[\int dx'\sqrt{-g^{(bk)}}
g^{\alpha\beta}{}_{\gamma'\delta'}(x,x')
\frac{d^2}{df^2}T^{\gamma'\delta'}(x';{\cal E}(f),{\cal C}(f))
\right]_{f=f(x)} f^{;\epsilon}(x)f_{;\epsilon}(x) 
\,. 
\nonumber \\ 
\label{eq:l_ein11'} 
\end{eqnarray}
The first term has a part of $O(\lambda^2)$ 
dependent on $(d/df){\cal C}$. 
However, by taking a smooth foliation, 
one can make $f^{;\epsilon}{}_{;\epsilon}(x)$ arbitrarily small, 
and we can eliminate this term. 
The second term also has a part dependent on $(d/df){\cal C}$, 
but it is of $O(\lambda^3)$ 
and can be taken as the source term for $h^{(1,2)\alpha\beta}$. 
Thus, the source term of (\ref{eq:l_ein11}) does not have 
an explicit dependence on the elapsed time of $(d/df){\cal C}$, 
and we avoid the Burke problem for $h^{(1,1)\alpha\beta}$. 

We next treat $h^{(2,0)\alpha\beta}$ 
by performing the adiabatic expansion of (\ref{eq:l_ein2}). 
The equation for $h^{(2,0)\alpha\beta}$ becomes 
\begin{eqnarray}
&& \frac{\lambda^2}{2}\left(
h^{(2,0)\alpha\beta;\gamma}{}_{;\gamma}
+2R^\alpha{}_\gamma{}^\beta{}_\delta h^{(2,0)\gamma\delta}
\right) 
\nonumber \\ && \qquad 
~=~ -\left[G^{\alpha\beta}\right]^{(2)}
(\lambda h^{(1,0)},\lambda h^{(1,0)})
+\Lambda^{\alpha\beta}(\lambda h^{(1,0)}) 
+O(\lambda^3)\,. \label{eq:l_ein20} 
\end{eqnarray}
The first term on the RHS of (\ref{eq:l_ein20}) has 
a term dependent on $(d/df){\cal C}$, 
but it is of $O(\lambda^3)$. 
For the second term on the RHS, instead of using (\ref{eq:l_mp10}), 
we consider the equation for $\Lambda^{\alpha\beta}(\lambda h^{(1)})$. 
By a simple manipulation, we obtain 
\begin{eqnarray}
&& \left[\Lambda^{\alpha\beta;\gamma}{}_{;\gamma}
+2R^\alpha{}_\gamma{}^\beta{}_\delta\Lambda^{\gamma\delta}\right]
(\lambda h^{(1)}) 
\nonumber \\ && \qquad 
~=~ -8\pi \left(T^{\alpha\gamma}{}_{;\gamma}{}^{;\beta}
+T^{\beta\gamma}{}_{;\gamma}{}^{;\alpha}
-\frac{1}{2}g^{(bk)\alpha\beta}T^{\gamma\delta}{}_{;\gamma\delta}
\right) 
\,. \label{eq:l_gau1}
\end{eqnarray}
Because we consider the stress-energy tensor of the point particles, 
we can evaluate the RHS of (\ref{eq:l_gau1}) explicitly, 
and we have 
\begin{eqnarray}
&& T^{\alpha\gamma}{}_{;\gamma}{}^{;\beta}
+T^{\beta\gamma}{}_{;\gamma}{}^{;\alpha}
-\frac{1}{2}g^{(bk)\alpha\beta}T^{\gamma\delta}{}_{;\gamma\delta}
\nonumber \\ 
&=& \sum_i m_i \int d\tau_i 
\left(\frac{D^2z^\alpha_i}{d\tau_i^2}\nabla^\beta
+\frac{D^2z^\beta_i}{d\tau_i^2}\nabla^\alpha
-\frac{1}{2}g^{(bk)\alpha\beta}
\frac{D^2z^\gamma_i}{d\tau_i^2}\nabla^\gamma\right)
\frac{\delta^{(4)}(x-z_i(\tau_i))}{\sqrt{-g^{(bk)}(x)}} \,, 
\nonumber \\ 
\end{eqnarray}
where $m_i$, $\tau_i$ and $z^\alpha_i$ are 
the mass, the proper time and the orbital coordinates 
of the $i$th particle. 
Here, $(D^2z^\alpha_i/d\tau_i^2)$ is the $4$-acceleration 
of the $i$th particle, 
and it depends on $(d/dt){\cal E}$, not on $(d/dt){\cal C}$. 
We consider integration of (\ref{eq:l_gau1}) 
using the adiabatic expansion method in \S \ref{sec:amp}. 
The result is not exactly equal 
to $\Lambda^{\alpha\beta}(\lambda h^{(1,0)})$, 
because the procedure of the adiabatic expansion 
does not commute with the derivative. 
However, to leading order in $\lambda$, 
the adiabatic expansion and the derivative operation commute. 
Thus, up to $O(\lambda^2)$, 
$\Lambda^{\alpha\beta}(\lambda h^{(1,0)})$ 
does not contain $(d/dt){\cal C}$ in an explicit manner. 
Because the source term of (\ref{eq:l_ein20}) does not have 
explicit dependence on the elapsed time of $(d/df){\cal C}$, 
we avoid the Burke problem for $h^{(2,0)\alpha\beta}$. 

Thus, we see that the source terms 
for $h^{(1,1)\alpha\beta}$ and $h^{(2,0)\alpha\beta}$, 
the metric perturbations of $O(\lambda^2)$ 
with this specific gauge fixing term 
do not depend on $(d/dt){\cal C}$, 
which grows linearly in the elapsed time 
after we switch on the radiation reaction. 
Although the precise behavior may still depend 
on the detailed form of the background metric, 
these source terms induce metric perturbations 
that do not grow in time, 
and the perturbation expansion in $\lambda$ holds 
for a sufficiently long time, 
i.e. longer than the radiation reaction time.

\section{Summary and discussion} \label{sec:con}

In this paper, we have investigated the so-called gauge problem 
arising in the study of the gravitational radiation reaction problem 
to a multi-particle system. 
Because the radiation reaction time scale is longer 
than the dynamical time scale of the system, 
we consider the use of a perturbation method 
to study the effect of the radiation reaction. 

Due to the radiation reaction, 
the evolution of the system is dissipative. 
There are some special perturbation techniques 
that can be applied o study a dissipative system, 
for example, a damped harmonic oscillator. 
The key problem that arises when using the standard perturbation expansion 
is that the perturbation expansion does not remain valid 
as the system evolves. 
Because the dissipative effect accumulates in time, 
the effect of the perturbation could become dominant over the background, 
and therefore the perturbation calculation may lose 
its predictive ability. 
In the case of a gravitational perturbation, 
such a problem could result 
from an inappropriate choice of the gauge condition. 
In Sec.\ref{sec:pri}, we first pointed out this problem 
in the study of the radiation reaction given in Ref.\cite{adi}, 
in which we call it the ``Burke problem''. 
A crucial consequence of this problem is that 
the evolution of the system becomes non-adiabatic 
despite the fact that 
the prediction under the adiabatic approximation is widely accepted, 
because of the observational evidence of the Hulse-Taylor binary. 

We considered what condition must be imposed on the gauge 
in order to avoid this problem. 
For this purpose, we proposed a new metric perturbation scheme 
in \S \ref{sec:rmp}. 
The problem of the standard metric perturbation was 
discussed in Ref.\cite{adi}. 
Because the linearized Einstein operator 
is algebraically divergence free, 
if we simply chooose the stress-energy tensor 
as the source term of the linearized Einstein tensor, 
the stress-energy tensor must be conserved in the background, 
and we cannot consistently take into account 
the radiation reaction effect on the matter to leading order. 
By adding a gauge fixing term, we can avoid this problem. 

Here we comment that adding the gauge fixing term is not 
what we usually do in the black hole perturbation calculation, 
such as the Zerilli-Regge-Wheeler formalism.\cite{ZRW} 
The Zerilli-Regge-Wheeler formalism is a convenient method 
to calculate a linear metric perturbation 
of a Schwarzschild black hole. 
In this formalism, we do not add a gauge fixing term 
to the linearized Einstein equation. 
Instead, we explicitly fix 
some components of the metric perturbation. 
In this procedure, the stress-energy tensor 
as a source of the metric perturbation 
must satisfy the conservation law in the background, 
because we only have six components of the metric perturbation. 
We could define the Green function 
of the Zerilli-Regge-Wheeler formalism. 
However, it would contain only 
the propagation of the gravitational mode, 
because of the explicit gauge fixing. 
Hence, this Green function would not be sufficient, 
as a Green function including propagation of the vector and scalar modes 
is necessary in this new perturbation scheme. 
The gauge fixing term in this new perturbation scheme does not vanish, 
and one can easily see that it comes 
from the vector and scalar modes of the Green function, 
because it is induced by $T^{\alpha\beta}{}_{;\beta}$. 
Therefore, it is necessary to extend the Zerilli-Regge-Wheeler formalism 
in order to apply it to the new metric perturbation scheme. 
We shall discuss this problem elsewhere. 

Although one may consider 
the dissipative effect of the particles' motion 
to lower order in the metric perturbation 
using the new metric perturbation scheme, 
a systematic calculation 
of higher-order metric perturbations 
cannot be derived from lower-order ones. 
In \S \ref{sec:amp}, we propose 
another perturbation method to avoid this technical problem, 
which was originally discussed in Ref.\cite{adi}. 

Using this new metric perturbation scheme, 
we turned to the Burke problem 
with a specific choice of the gauge fixing term in \S \ref{sec:bur}. 
For a complete solution of the Burke problem, 
it is necessary to determine the behavior of every order 
in the perturbation expansion of the metric. 
However, this is a difficult task. 
In \S \ref{sec:bur}, 
we only computed the metric perturnation to second order. 
We further supposed that the Green function is well-behaved 
in the sense that, if the source term does not grow in time, 
the resulting field will not grow in time. 
Under this condition, we find that our choice of the gauge fixing term 
allows us to avoid the Burke problem.

\section*{Acknowledgement}

Y.M. thanks Prof. Kip S. Thorne, Prof. Clifford M. Will 
and Prof. Eanna Flanagan for fruitful discussions, 
and Prof. E. Sterl Phinney for encouragement. 

Y.M. was supported by the NASA Grant NAG5-12834 
and the NASA-ATP Grant NNG04GK98G. 



\end{document}